\documentclass[3p,times,twocolumn]{elsarticle}

\usepackage{ecrc}

\volume{00}

\firstpage{1}

\journalname{Nuclear Physics B Proceedings Supplement}

\runauth{}

\jid{nuphbp}

\jnltitlelogo{Nuclear Physics B Proceedings Supplement}

\usepackage{amssymb,amsmath}

\usepackage[figuresright]{rotating}

\begin{document}

\begin{frontmatter}



\dochead{}

\title{Radiatively-induced LFV Higgs Decays from Massive ISS Neutrinos}


\author[unizar]{E.~Arganda\corref{cor1}}
\cortext[cor1]{Talk given by E. Arganda at the 37$^{th}$ International Conference on High Energy Physics (ICHEP 2014).}
\ead{ernesto.arganda@unizar.es}
\author[ift]{M.~J.~Herrero}
\ead{maria.herrero@uam.es}
\author[ift]{X.~Marcano}
\ead{xabier.marcano@uam.es}
\author[ift]{C.~Weiland}
\ead{cedric.weiland@uam.es}
\address[unizar]{Departamento de F\'{\i}sica Te\'orica, Facultad de Ciencias,\\
Universidad de Zaragoza, E-50009 Zaragoza, Spain}
\address[ift]{Departamento de F\'{\i}sica Te\'orica and Instituto de F\'{\i}sica Te\'orica, IFT-UAM/CSIC,\\
Universidad Aut\'onoma de Madrid, Cantoblanco, 28049 Madrid, Spain}

\begin{abstract}
In the inverse seesaw model (ISS), the smallness of the neutrino masses is related to the smallness of a lepton number violating mass term whilst the seesaw scale is naturally close to the TeV scale, which allows for large effects in lepton flavor and universality violating observables. With the ongoing and planned measurements of the Higgs boson properties at the LHC, we found timely to investigate the possibility of having large lepton flavor violating Higgs decay (LFVHD) rates within the context of the ISS, considering the most generic case where three additional pairs of massive right-handed singlet neutrinos are added to the Standard Model particle content. We present a full one-loop computation of the LFVHD rates and analyze in full detail the predictions as functions of the various relevant ISS parameters, which are required to be compatible with the present neutrino data and the present experimental bounds for the three LFV radiative decays, and also consistent with other constraints, like perturbativity of the neutrino Yukawa couplings. At the end, we conclude on the maximum allowed LFVHD rates within the ISS, which may reach maximal values of order $10^{-5}$ for the $H \to e \bar \tau$ and $H \to \mu \bar \tau$ channels, close to the expected future LHC sensitivities.
\end{abstract}


\end{frontmatter}


\section{Introduction}
\label{intro}

The fact of considering the discovered scalar particle at the CERN-LHC as the Higgs particle of the Standard Model (SM), with very similar properties and a measured mass of $m_h^{\rm ATLAS}=125.5\pm 0.6 $ GeV~\cite{Aad:2013wqa} and $m_h^{\rm CMS}=125.7\pm 0.4 $ GeV~\cite{Chatrchyan:2013lba}, has at present reached a broad consensus.

On the other hand, there is also a major consensus about the fact that the SM must be modified to include the neutrino masses and oscillations in agreement with present data~\cite{GonzalezGarcia:2012sz}, which are nowadays quite impressive and call for an explanation from a theoretical framework beyond the SM. In that sense, we consider here one of the simplest and most appealing extensions of the SM, the Inverse Seesaw Model (ISS)~\cite{Mohapatra:1986aw,Mohapatra:1986bd,Bernabeu:1987gr}, which extends the SM particle content by adding pairs of right-handed (RH) neutrinos with opposite lepton number. The seesaw mechanism that produces the small light physical neutrino masses in the ISS is associated to the smallness of the Majorana mass model parameters and it allows for large Yukawa neutrino couplings while having at the same time moderately heavy RH neutrino masses at the ${\cal O}({\rm TeV})$ energies which are 
reachable at the present colliders, like the LHC. In addition, these RH neutrinos can produce non-negligible contributions to processes with Lepton Flavor Violation (LFV) via radiative corrections that are mediated by the sizable neutrino Yukawa couplings, therefore leading to some hint of these rare processes, which are totally absent in the SM .

Combining Higgs physics with LFV, we study, within the ISS context with three extra pairs of RH neutrinos, the Higgs decays into lepton-antilepton pairs $H \to l_k\bar l_m$ with $k \neq m$ (LFVHD), which are being currently explored at the LHC~\cite{CMS:2014hha}. We refer the reader to our main article~\cite{Arganda:2014dta} for more information and details about the full one-loop computation of the LFV partial decay widths, the complete set of references and our full numerical results.

\section{The Inverse Seesaw Model}
\label{ISSmodel}

The ISS supplements the SM with pairs of RH neutrinos, denoted here by $\nu_R$ and $X$, with opposite lepton number. We consider a generic model containing three pairs of fermionic singlets, extending the SM Lagrangian with the following neutrino Yukawa interactions and mass terms:
\begin{equation}
 \label{LagrangianISS}
 \mathcal{L}_\mathrm{ISS} = - Y^{ij}_\nu \overline{L_{i}} \widetilde{\Phi} \nu_{Rj} - M_R^{ij} \overline{\nu_{Ri}^C} X_j - \frac{1}{2} \mu_{X}^{ij} \overline{X_{i}^C} X_{j} + h.c.\,,
\end{equation}
where $L$ is the SM lepton doublet, $\Phi$ is the SM Higgs doublet, $\widetilde{\Phi}=\imath \sigma_2 \Phi^*$, with $\sigma_2$ the corresponding Pauli matrix, $Y_\nu$ is the $3\times3$ neutrino Yukawa coupling matrix, $M_R$ is a lepton number conserving complex $3\times3$ mass matrix, and $\mu_X$ is a Majorana complex $3\times3$ symmetric mass matrix that violates lepton number conservation by two units.
After electroweak symmetry breaking, the $9\times 9$ neutrino mass matrix reads, in the electroweak interaction basis $(\nu_L^C\,,\;\nu_R\,,\;X)$,
\begin{equation}
\label{ISSmatrix}
 M_{\mathrm{ISS}}=\left(\begin{array}{c c c} 0 & m_D & 0 \\ m_D^T & 0 & M_R \\ 0 & M_R^T & \mu_X \end{array}\right)\,,
\end{equation}
with the $3\times3$ Dirac mass matrix given by $m_D=Y_\nu \langle \Phi\rangle$, and the Higgs vacuum expectation value taken to be $\langle \Phi\rangle=v = 174\,\mathrm{GeV}$.

In the one generation case, which allows us to illustrate more simply the dependence on the seesaw parameters, there would be just three ISS model parameters, $M_R$, $\mu_X$ and $Y_\nu$, and there would be just three physical eigenstates: one light $\nu$ and two heavy $N_1$ and $N_2$. In the limit $\mu_X \ll m_D, M_R$, the mass eigenvalues are given by:
\begin{align}
 m_\nu &= \frac{m_{D}^2}{m_{D}^2+M_{R}^2} \mu_X\,\label{mnu},\\
 m_{N_1,N_2} &= \pm \sqrt{M_{R}^2+m_{D}^2} + \frac{M_{R}^2 \mu_X}{2 (m_{D}^2+M_{R}^2)}\,,\label{mN}
\end{align}
with the light neutrino mass $m_\nu$ being proportional to $\mu_X$, thus making it naturally small, and the two heavy masses $m_{N_1,N_2}$ being close to each other. As a consequence in this $\mu_X \ll m_D, M_R$ limit, these two nearly degenerate heavy neutrinos combine to form a pseudo-Dirac fermion. A similar pattern of neutrino mass eigenvalues occurs in the three generation case, with one light and two nearly degenerate heavy neutrinos per generation.

We will consider first the simplest ISS scenarios with diagonal $\mu_X$ and $M_R$ matrices (case I). In order to implement easily the compatibility with present neutrino data in these scenarios, we use here the helpful Casas-Ibarra parametrization~\cite{Casas:2001sr} that can be directly applied to the ISS case, giving:
\begin{eqnarray}
 m_D^T &=& V^\dagger \mathrm{diag}(\sqrt{M_1}\,,\sqrt{M_2}\,,\sqrt{M_3})\; R\; \nonumber \\ 
 & \times & \mathrm{diag}(\sqrt{m_{\nu_1}}\,, \sqrt{m_{\nu_2}}\,, \sqrt{m_{\nu_3}}) U^\dagger_{\rm PMNS}\,,
\label{CasasIbarraISS}
\end{eqnarray}
where $V$ is a unitary matrix that diagonalizes $M=M_R \mu_X^{-1} M_R^T$ according to $ M=V^\dagger \mathrm{diag}(M_1\,, M_2\,, M_3) V^*$ and $R$ is a complex orthogonal matrix that can be written in terms of three arbitrary complex angles $\theta_{1,2,3}$. The input ISS parameters that have to be fixed in this case I are the following: $m_{\nu_{1,2,3}}$, $\mu_{X_{1,2,3}}$, $M_{R_{1,2,3}}$, $\theta_{1,2,3}$ and the entries of the $U_{\rm PMNS}$ matrix.

On the other hand, given the interesting possibility of decoupling the low energy neutrino physics from the LFV physics in this ISS model by the proper choice of the input parameters, we will take into account specific ISS scenarios with non-diagonal $\mu_X$ while keeping diagonal $M_R$ (case II) which can provide the largest LFV Higgs decay rates. Once some specific inputs are provided for $Y_\nu$ and $M_R$, the proper $\mu_X$ matrix that ensures the agreement between low energy neutrino predictions and data 
can be easily obtained by solving ISS equations, which leads to:
\begin{equation}\label{muXtexture}
\mu_X=M_R^T ~m_D^{-1}~ U_{\rm PMNS}^* m_\nu U_{\rm PMNS}^\dagger~ {m_D^T}^{-1} M_R
\end{equation}
with $m_D= v Y_\nu$ and $m_\nu =\mathrm{diag}(m_{\nu_1}\,, m_{\nu_2}\,, m_{\nu_3})$. It should be noted that for a generic $Y_\nu$ texture, this $\mu_X$ will be in general non-diagonal. Therefore, the most relevant input ISS parameters in the case II are $Y_\nu$ and $M_R$ (in addition to the input parameters that are relevant for low energy neutrino physics, $m_\nu$ and $U_{PMNS}$).

\section{Remarks on the LFVHD Computation}
\label{computation-remarks}

In our one-loop computation of the LFV rates we work in the mass basis for all the particles involved, with diagonal charged leptons, and take into account the contributions from all the nine physical neutrinos. As for the gauge choice, we choose the Feynman-t'Hooft gauge. The full set of contributing one-loop diagrams can be found in~\cite{Arganda:2014dta}, and we have adapted the complete one-loop formulas for the $\Gamma(H \to l_k\bar l_m)$ partial decay width, taken from~\cite{Arganda:2004bz}, to the ISS case. We have focused on the decays $H\to \mu\bar\tau,e\bar\tau,e\bar\mu$ and have not considered their related $CP$ conjugate decays $H\to\tau\bar\mu,\tau\bar e,\mu \bar e$ which, in the presence of complex phases, could lead to different rates. All the formulas for the LFVHD have been implemented into our private {\tt Mathematica} code. For the numerical predictions of the BR$(H \to {l_k} \bar{l}_m)$ rates, we use
$m_H=126\,{\rm GeV}$ and its corresponding SM total width is computed with {\tt FeynHiggs}~\cite{Heinemeyer:1998yj,Heinemeyer:1998np,Degrassi:2002fi} including two-loop corrections.

In order to ensure the validity of the Casas-Ibarra parametrization, we have imposed that the error on the light neutrino masses estimated with it, meaning the differences between the input $m_{\nu_{1,2,3}}$ and the output $m_{n_{1,2,3}}$ masses, was below $10\%$ and that the $9 \times 9$ rotation matrix exhibited the required unitarity property. Furthermore, since a given set of input parameters can generate arbitrarily large Yukawa couplings, we enforce their perturbativity by setting an upper limit on the entries of the neutrino Yukawa coupling matrix, given by
\begin{equation}\label{Yuk-pert}
\frac{|Y_{ij}|^2}{4\pi}<1.5 \,,
\end{equation}
for $i,j=1,2,3$.

At the same time that we analyze the LFVHD, we also compute the one-loop $l_m \to l_k \gamma$ decay rates within this same ISS framework (using the analytical formulas provided in~\cite{Ilakovac:1994kj} and~\cite{Deppisch:2004fa}) and for the same input parameters, and check that these radiative decay rates are compatible with their present experimental $90\%$ CL upper bounds:
\begin{align}
{\rm BR}(\mu\to e\gamma)&\leq 5.7\times 10^{-13}\text{\cite{Adam:2013mnn}}\label{MUEGmax}\,,\\
{\rm BR}(\tau\to e\gamma)&\leq 3.3\times 10^{-8}~\text{\cite{Aubert:2009ag}}\label{TAUEGmax}\,,\\
{\rm BR}(\tau\to \mu\gamma)&\leq 4.4\times 10^{-8}~\text{\cite{Aubert:2009ag}}\label{TAUMUGmax}\,.
\end{align}

Regarding the Higgs total width, we focus on the scenario where the new fermionic singlets have a mass above $200\,\mathrm{GeV}$, thus they do not contribute to new invisible decays. To avoid potential constraints from lepton electric dipole moments, we assume that all mass matrices are real, as well as the PMNS matrix.
Additional constraints might also arise from lepton universality tests~\cite{Abada:2012mc, Abada:2013aba}. Nevertheless, in the scenario that we consider where the sterile neutrinos are heavier than the Higgs boson, points that would be excluded by lepton universality tests are already excluded by $\mu \rightarrow e \gamma$. 
At the end, we found that the most stringent constraints for our study are by far $\mu \rightarrow e \gamma$ and the Yukawa coupling perturbativity limit of Eq.~(\ref{Yuk-pert}).

\section{LFV Rates in the Inverse Seesaw: Case I}
\label{resultsI}

We study first the LFV rates as functions of the most relevant ISS parameters within the case I, trying to localize the areas of the parameter space where the LFVHD can be both large and respect the constraints on the LFV radiative decays. The results will be presented in two generically different scenarios for the heavy neutrinos: 1) the case of (nearly) degenerate heavy neutrinos, and 2) the case of hierarchical heavy neutrinos.

\begin{figure}[t!]
\begin{center}
\begin{tabular}{c}
\includegraphics[width=0.45\textwidth]{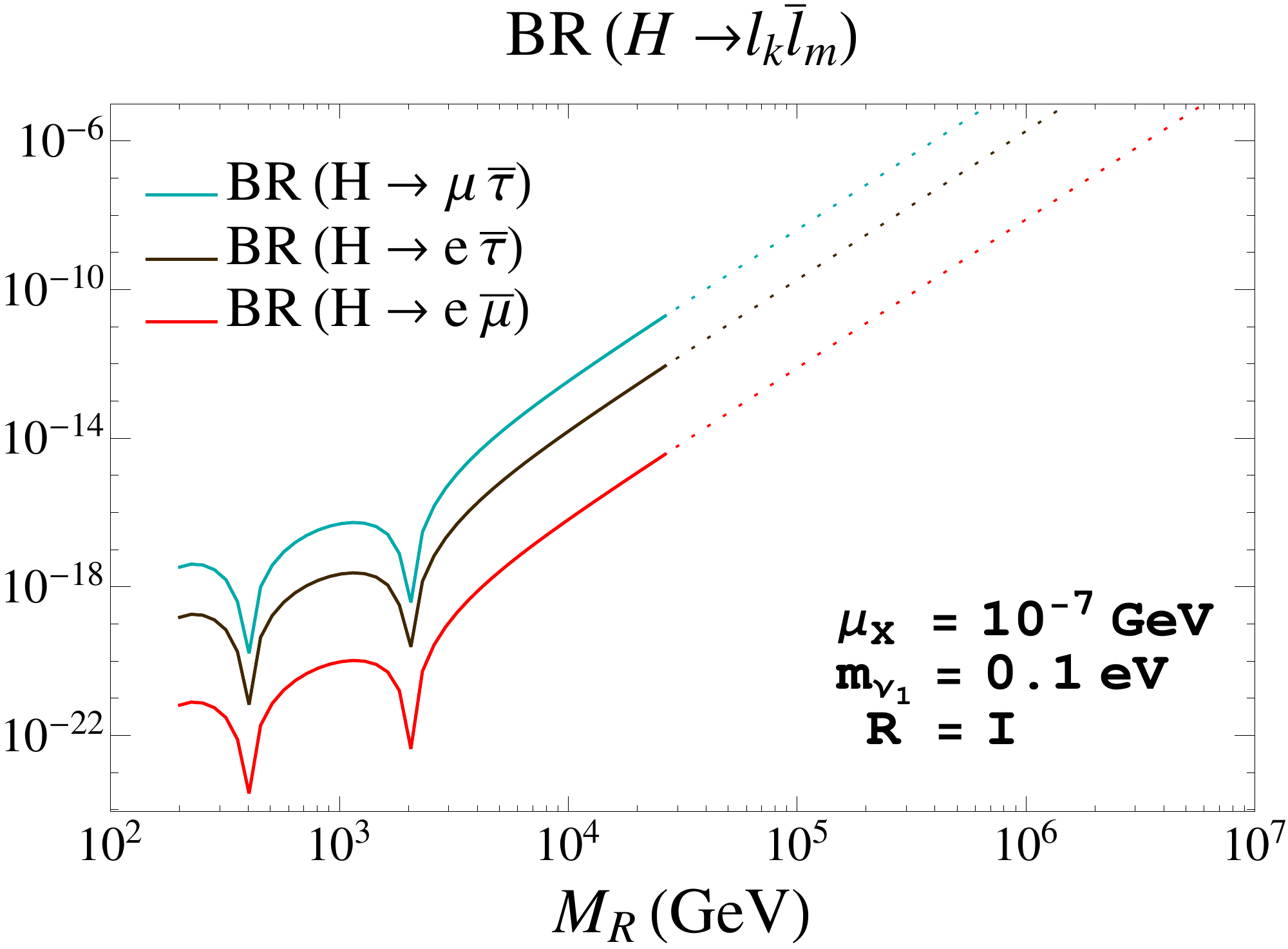} \\
\includegraphics[width=0.45\textwidth]{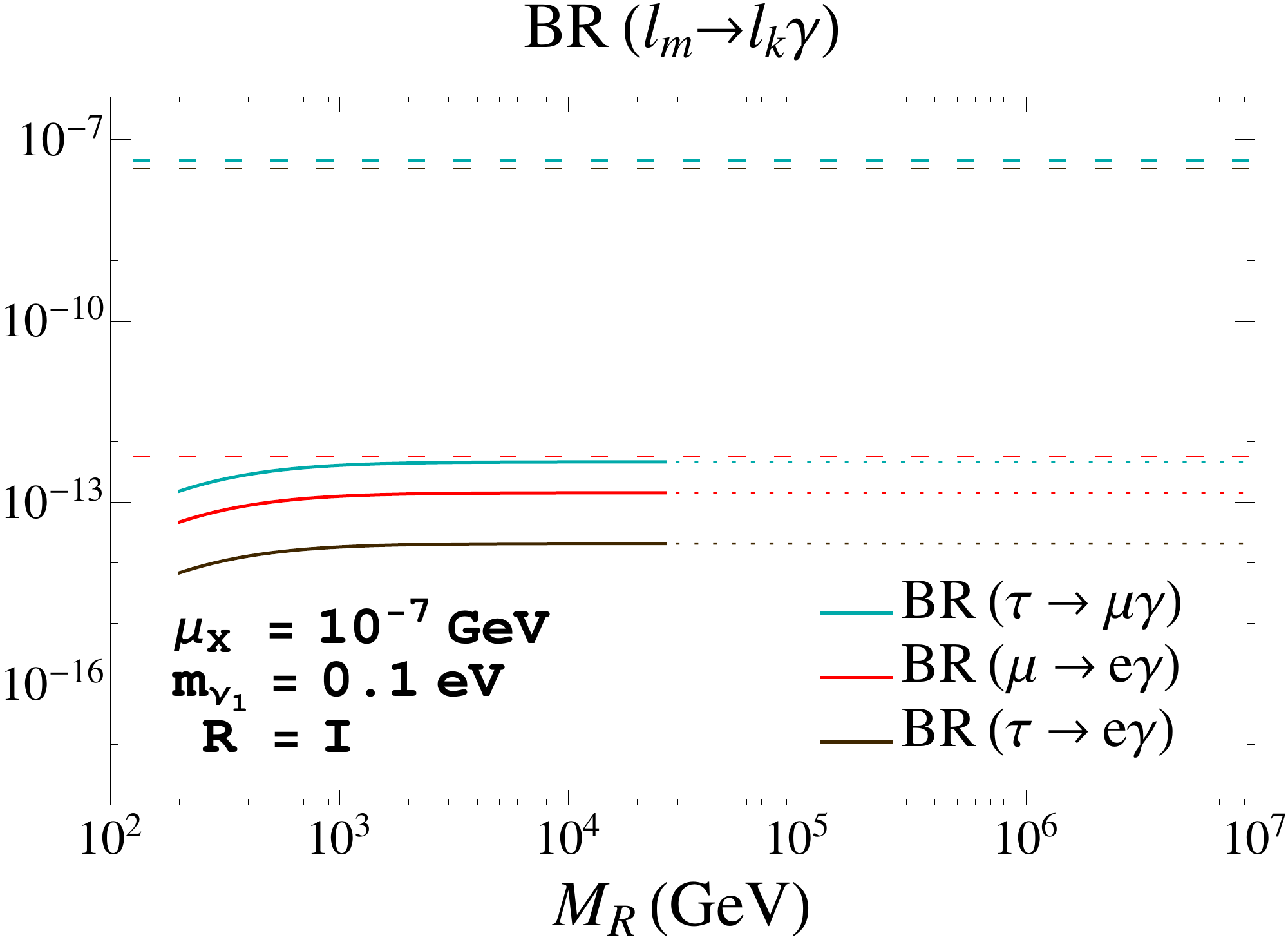}
\end{tabular}
\caption{Predictions for the LFV decay rates as functions of $M_R$ in the degenerate heavy neutrinos case. The dotted lines in both panels indicate non-perturbative neutrino Yukawa couplings. The horizontal dashed lines in the lower panel are the present ($90\%$ CL) upper bounds on the radiative decays: BR$(\tau \to \mu \gamma)<4.4 \times 10^{-8}$~\cite{Aubert:2009ag} (blue line),
 BR$(\tau \to e \gamma)<3.3 \times 10^{-8}$~\cite{Aubert:2009ag} (dark brown line), BR$(\mu \to e \gamma)< 5.7 \times 10^{-13}$~\cite{Adam:2013mnn} (red line).
}\label{ALL_LFVdecays}
\end{center}
\end{figure}

The case of (nearly) degenerate heavy neutrinos is implemented here by choosing degenerate entries in $M_{R_i} = M_R$ and $\mu_{X_i} = \mu_X$ ($i = 1, 2, 3$). First we show in Fig.~\ref{ALL_LFVdecays} the results for all LFV rates as functions of the common RH neutrino mass parameter $M_R$ for all LFVHD channels (upper panel) and for the LFV radiative decay channels (lower panel). As expected, we find that the largest LFVHD rates are for BR($H \to \mu\bar \tau$) and the largest radiative decay rates are for BR($\tau \to \mu \gamma$). We also see that, for this particular choice of input parameters, all the predictions for the LFVHD are allowed by the present experimental upper bounds on the three radiative decays for all explored values of $M_R$ in the interval $(200, 10^7)\, {\rm GeV}$. Besides, it shows clearly that the most constraining radiative decay at present is by far the $\mu \to e \gamma$ decay. This is so in all the cases 
explored in this work, so whenever we wish to conclude on the allowed LFVHD rates we will focus mainly on this radiative channel. 

Regarding the $M_R$ dependence shown in Fig.~\ref{ALL_LFVdecays}, we clearly see that the LFVHD rates grow faster with $M_R$ than the radiative decays which indeed tend to a constant value for $M_R$ above $\sim 10^3$ GeV. In fact, the LFVHD rates can reach quite sizable values in the large $M_R$ region of these plots, and yet stay allowed by the constraints on the radiative decays. For instance, BR$(H \to \mu\bar \tau) \sim 10^{-6}$ for $M_R = 4\times 10^5$ GeV. However, our requirement of perturbativity for the neutrino Yukawa coupling entries does not allow for such large $M_R$ values leading to too large $Y_\nu$ values in the framework of our parametrization of Eq.~(\ref{CasasIbarraISS}). Indeed, the exclusion region for $M_R$ from perturbativity of $Y_\nu$ forbids these large $M_R$ values. For the specific input parameter values of Fig.~\ref{ALL_LFVdecays}, the forbidden values are for $M_R$ above $3 \times 10^4$ GeV, and this leads to maximum allowed values of BR$(H \to \mu\bar \tau) \sim 2 \times10^{-11}$, BR$(H \to e \bar \tau) \sim 10^{-12}$ and BR$(H \to e \bar \mu) \sim 5 \times 10^{-15}$.

\begin{figure}[t!]
\begin{center}
\begin{tabular}{c}
\includegraphics[width=0.45\textwidth]{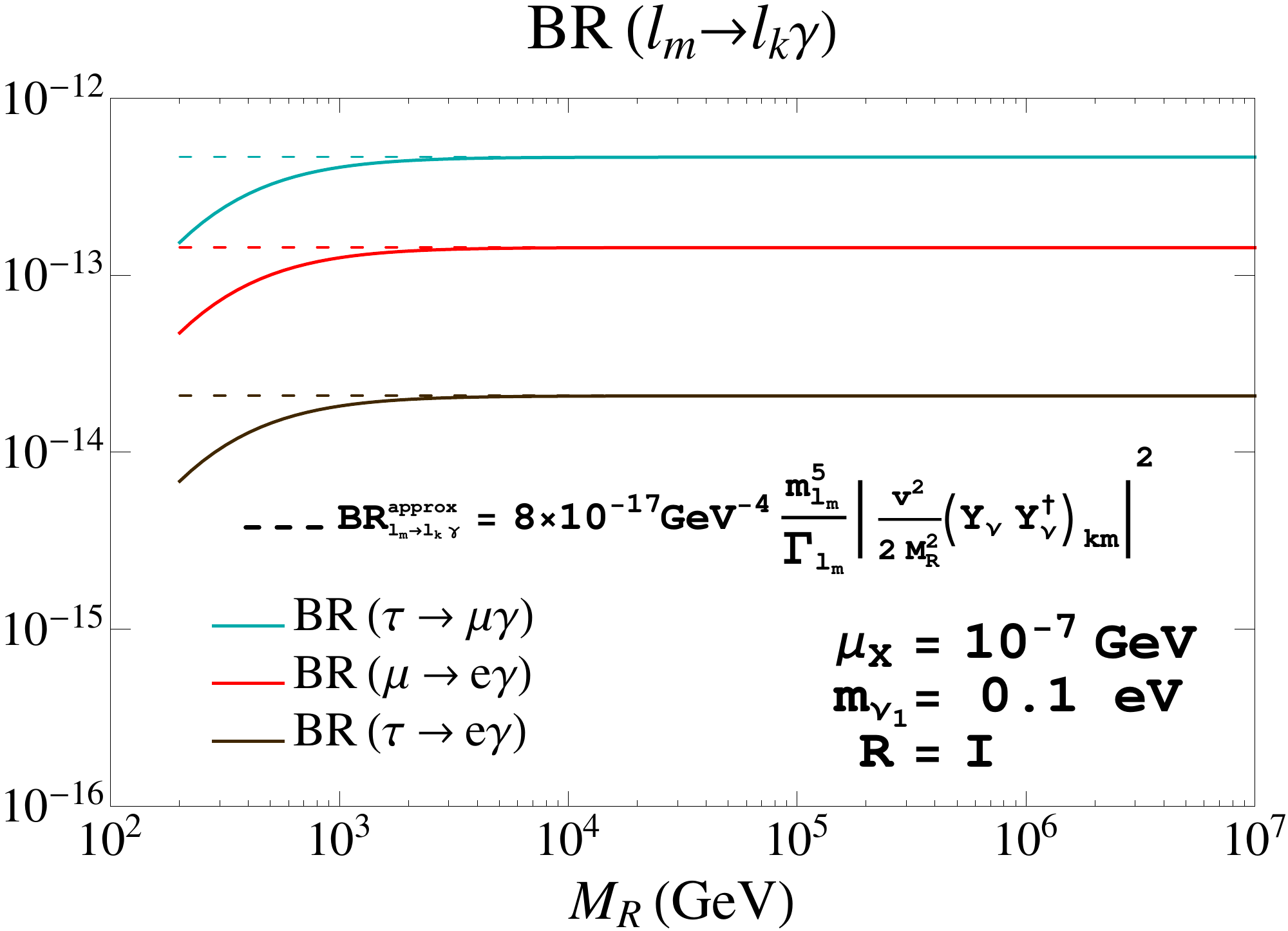}
\end{tabular}
\caption{Comparison of the full one-loop and approximate rates for the radiative decays $l_m \to l_k \gamma$ in the degenerate heavy neutrinos case. Dotted lines indicate non-perturbative neutrino Yukawa couplings. The other input parameters are set to $\mu_X= 10^{-7} \, {\rm GeV}$, 
$m_{\nu_1}=0.1 \, {\rm eV}$ and $R=I$.}\label{approxrad}
\end{center}
\end{figure}

The qualitatively different functional behavior with $M_R$ of the LFVHD and the radiative rates shown by Fig.~\ref{ALL_LFVdecays} is an interesting feature that we wish to explore further.
As it is clearly illustrated in Fig.~\ref{approxrad}, the radiative decay rates can be well approximated for large $M_R$ by a simple function of $|(Y_\nu Y_\nu^\dagger)_{km}|^2$ given by:
\begin{equation}
{\rm BR}^{\rm approx}_{l_m \to l_k \gamma}=8 \times 10^{-17} \frac{m_{l_m}^5({\rm GeV}^5)}{\Gamma_{l_m}{\rm (GeV)}} 
\bigg|\frac{v^2}{2M_R^2}(Y_\nu Y_\nu^\dagger)_{km}\bigg|^2,
\label{approxformula} 
\end{equation} 
which provides very close predictions to the exact rates for $M_R>10^3$ GeV. Then we can understand the final constant behavior of all the radiative decay rates with $M_R$, since the $|(Y_\nu Y_\nu^\dagger)_{km}|^2$ elements grow with $M_R$ approximately as $M_R^4$ in the parametrization used here of Eq.~(\ref{CasasIbarraISS}).  
This simple behavior with $M_R$ is certainly not the case of the LFVHD rates, and we conclude that these do not follow this same behavior with $|(Y_\nu Y_\nu^\dagger)_{km}|^2$.

\begin{figure}[t!] 
\begin{center}
\includegraphics[width=.47\textwidth]{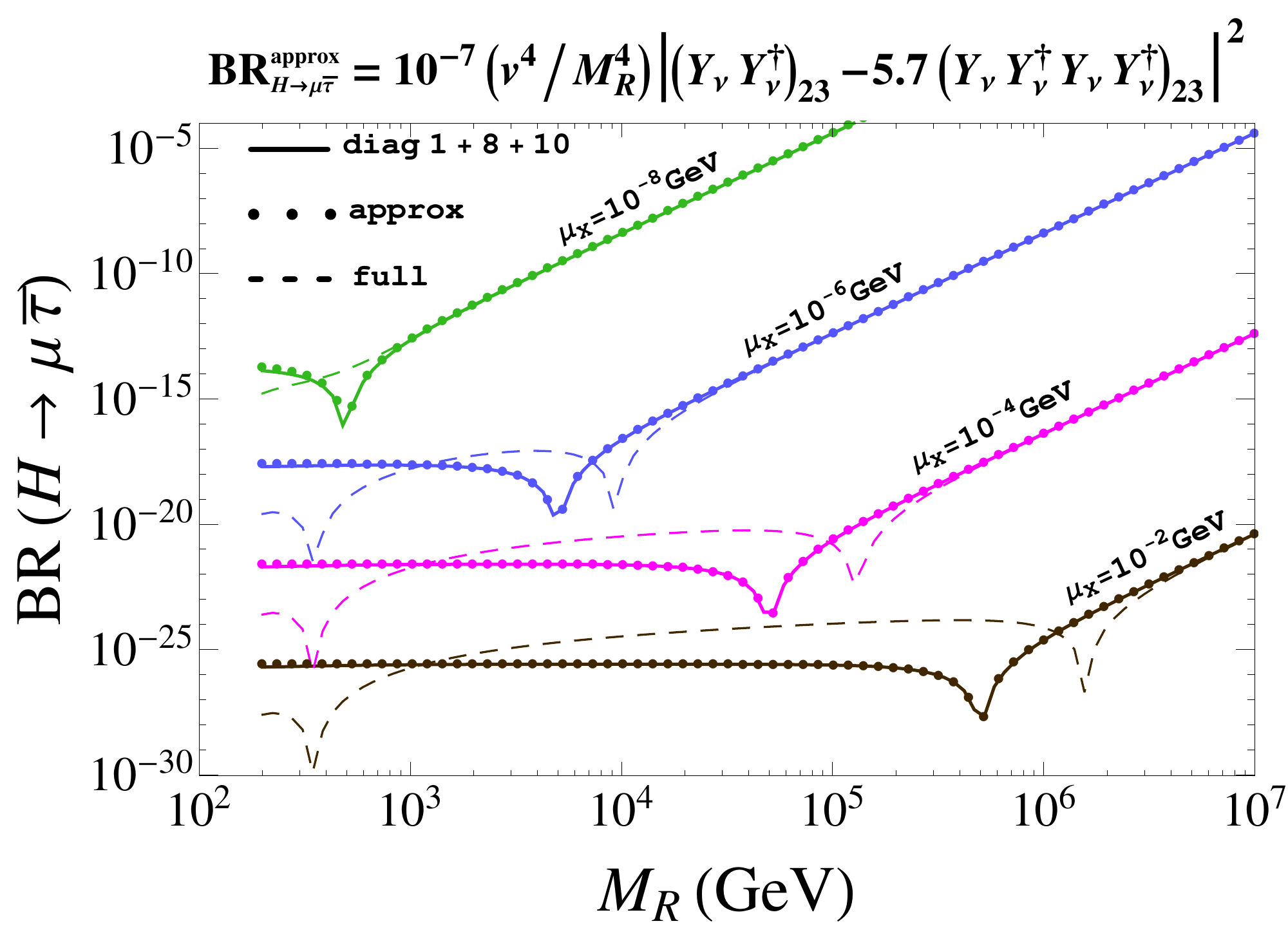} 
\caption{Comparison between the predicted rates for BR($H\to\mu\overline{\tau}$) taking: 1) the full one-loop formulas (dashed lines); 2) only the contributions from dominant diagrams (solid lines), and 3) the approximate formula of Eq.~(\ref{FIThtaumu}) (dotted lines).}
\label{fitLFVHD}
\end{center}
\end{figure}

In order to understand the different functional behavior of LFVHD rates with $M_R$, we have tried to find an approximate simple formula that could explain the main features of these rates. As we have already said, in contrast to what we have seen for the LFV radiative decays in Eq.~(\ref{approxformula}), a simple functional dependence being proportional to $|(Y_\nu Y_\nu^\dagger)_{23}|^2$ is not enough to describe our results for the BR($H \to \mu\overline\tau$) rates.
Considering that, in the region where the Yukawa couplings are large, we have looked for a simple expression that could fit properly the contributions from the dominant diagrams. From this fit we have found the following approximate formula:
\begin{equation}\label{FIThtaumu}
{\rm BR}^{\rm approx}_{H\to\mu\bar\tau}=10^{-7}\frac{v^4}{M_R^4}~\Big|(Y_\nu Y_\nu^\dagger)_{23}-5.7(Y_\nu Y_\nu^\dagger Y_\nu Y_\nu^\dagger)_{23}\Big|^2,
\end{equation}
which turns out to work reasonably well. In Fig.~\ref{fitLFVHD} we show   
the predicted rates of BR($H\to\mu\overline{\tau}$) with 1) the full one-loop formulas; 2) taking just the contributions from dominant diagrams, and 3) using Eq.~(\ref{FIThtaumu}) (dotted lines). We see clearly that this Eq.~(\ref{FIThtaumu}) reproduces extremely well the contributions from dominant diagrams and approximates reasonably well the full rates. The approximation is pretty good indeed for the $M_R$ region above the dips. The change of functional behavior with $M_R$ in the two different $M_R$ regions, from nearly flat with $M_R$ in the approximate result to fast growing as $\sim M_R^4$, also gives a reasonable approach to the full result, as well as the appearance of dips. The location of the dips is however not so accurately described by the approximate formula, since in the region where the cancellation among the dominant diagrams takes place the other diagrams (not considered in the fit) also contribute. Overall we find the approximate formula given by Eq.~(\ref{FIThtaumu}) very useful for generic estimates in the ISS, which could be also applied to other parametrizations of the neutrino Yukawa couplings.  

\begin{figure}[t!] 
\begin{center}
\begin{tabular}{c}
\includegraphics[width=0.45\textwidth]{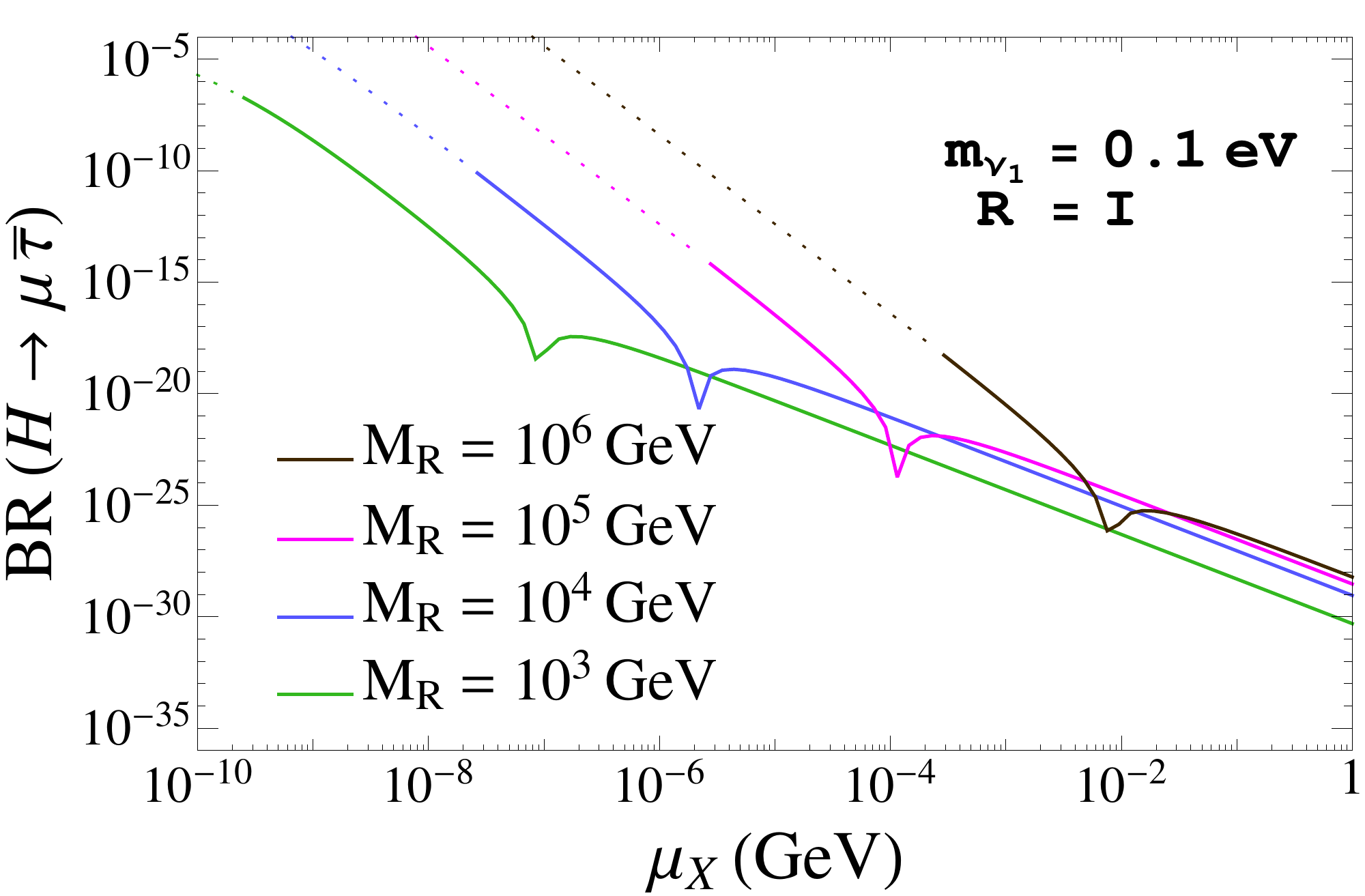} \\
\includegraphics[width=0.45\textwidth]{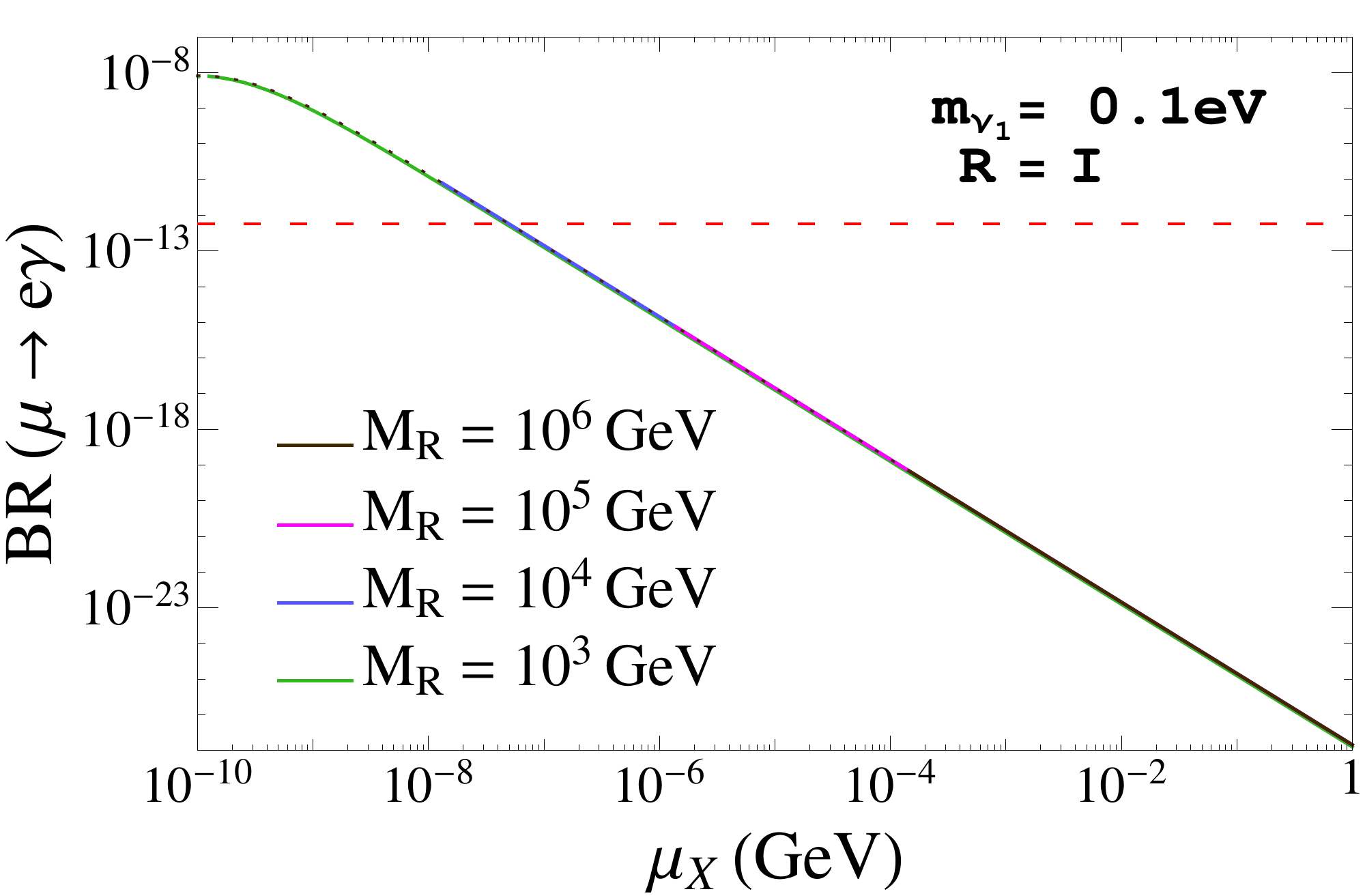}
\end{tabular}
\caption{Branching ratios of $H \to \mu\overline\tau$ (upper panel) and $\mu \to e \gamma$ (lower panel) as functions of $\mu_X$ for different values of $M_R=(10^{6},10^{5},10^{4},10^{3})$ GeV from top to bottom. In both panels, $m_{\nu_1} = 0.1$ eV and $R = I$. The horizontal red dashed line denotes the current experimental upper bound for $\mu \to e \gamma$, BR($\mu \to e \gamma$) $< 5.7 \times 10^{-13}$~\cite{Adam:2013mnn}. Dotted lines represent non-perturbative neutrino Yukawa couplings.}\label{BRs_muX_degenerate}
\end{center}
\end{figure}

Next we study the dependence of the LFV rates on $\mu_X$. The behavior of BR($H \to \mu\overline\tau$) and BR($\mu \to e \gamma$) as functions of $\mu_X$, for several values of $M_R$, $m_{\nu_1} = 0.1$ eV and $R = I$, are displayed in Fig.~\ref{BRs_muX_degenerate}. Both LFV rates decrease as $\mu_X$ grows; however, the functional dependence is not the same. The LFV radiative decay rates decrease as $\mu_X^{-2}$, in agreement with the approximate expression (\ref{approxformula}), while the LFVHD rates go as $\mu_X^{-4}$ when the Yukawa couplings are large. For a fixed value of $\mu_X$, the larger $M_R$ is, the larger BR($H \to \mu\overline\tau$) can be reached, while the same prediction for BR($\mu \to e \gamma$) is obtained for any value of $M_R$. We have already learnt this independence of the LFV radiative decays on $M_R$ from the previous figure, which can be easily confirmed on the lower panel of Fig.~\ref{BRs_muX_degenerate}, where all the lines for different values of $M_R$ are superimposed. We observe again the existence of dips in the upper panel of Fig.~\ref{BRs_muX_degenerate}. We see in this figure that the smallest value of $\mu_X$ allowed by the BR($\mu \to e \gamma$) upper bound is $\mu_X\sim 5\times10^{-8}$ GeV, which is directly translated to a maximum allowed value of BR($H \to \mu\overline\tau$) $\sim 10^{-11}$, for $M_R = 10^4$ GeV.

\begin{figure}[t!] 
\begin{center}
\begin{tabular}{cc}
\hspace{-0.5cm}\includegraphics[width=.47\textwidth]{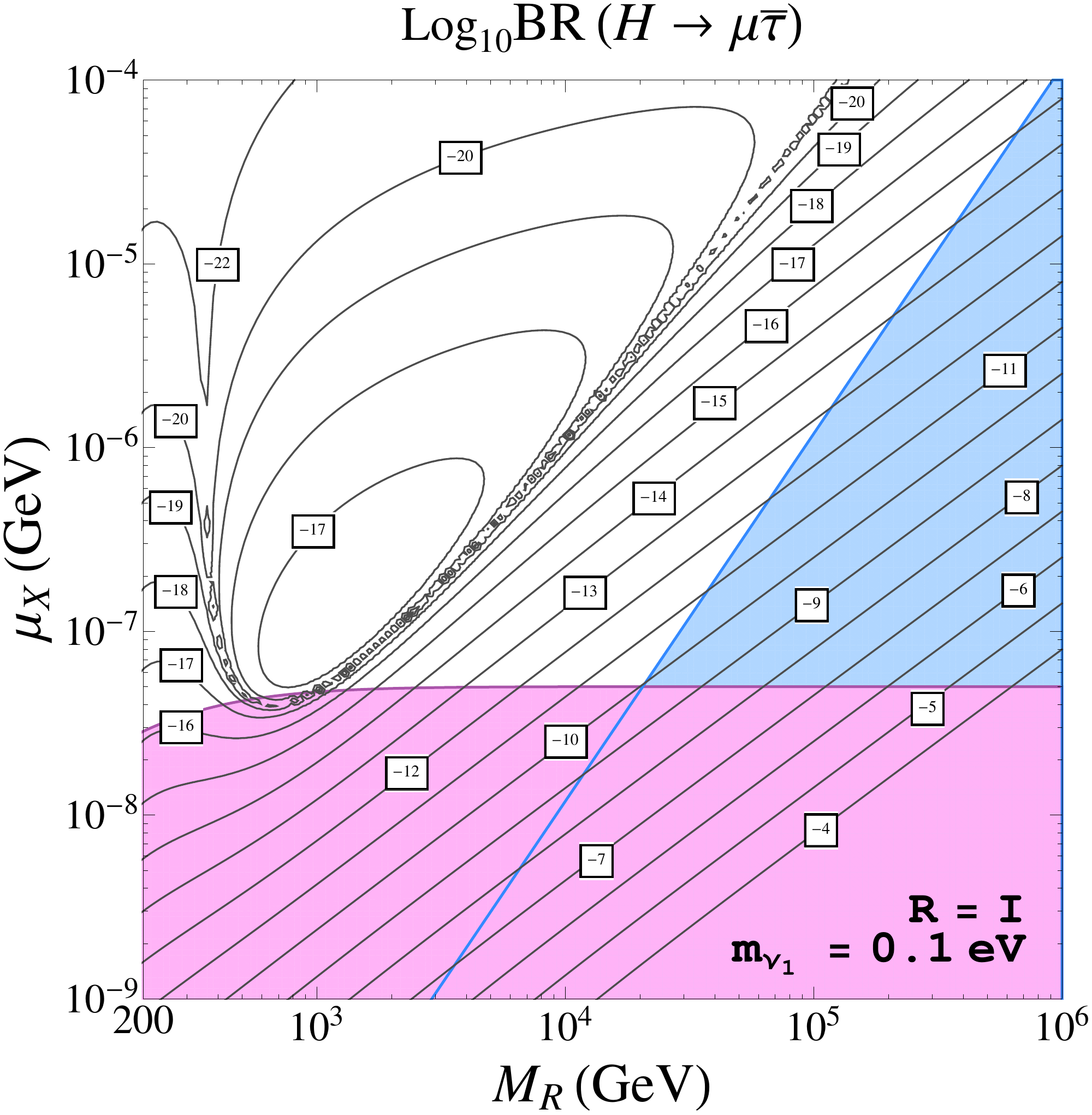} &
\end{tabular}
\caption{Contour lines of BR($H \to \mu \bar \tau$) in the $(M_R,\mu_X)$ plane for $R=I$ and $m_{\nu_1}=0.1\,{\rm eV}$. The shaded pink area is excluded by the upper bound on BR$(\mu \to e \gamma)$. The shaded blue area is excluded by the perturbativity requirement of the neutrino Yukawa couplings.}
\label{ContourPlot_degenerate}
\end{center}
\end{figure}

Once we have studied the behavior of all the LFV observables considered here with the most relevant parameters, we next present the results for the maximum allowed LFVHD rates in the case of heavy degenerate neutrinos. The plot in Fig.~\ref{ContourPlot_degenerate} shows the contour lines of BR($H \to \mu \bar \tau$) in the $(M_R,\mu_X)$ plane for $R=I$ and $m_{\nu_1}=0.1\,{\rm eV}$. These contour lines summarize the previously learnt behavior with $M_R$ and $\mu_X$, which lead to the largest values for the LFVHD rates in the bottom right-hand corner of the plot, i.e. at large $M_R$ and small $\mu_X$. We also notice the appearance of dips in the $(M_R,\mu_X)$ plane which correspond to the dips commented before in the previous figures. The most important 
conclusion from this contour plot is that the maximum allowed LFVHD rate for this simple hypothesis of diagonal and degenerate $\mu_X$ and $M_R$ matrices is approximately BR$(H \to \mu \bar \tau)\sim 10^{-10}$ and it is found for $M_R\sim 2\times 10^4\,{\rm GeV}$ and $\mu_X \sim  5\times 10^{-8} \, {\rm GeV}$. We have found similar conclusions for BR$(H \to e \bar \tau)$.  

The case of hierarchical heavy neutrinos refers here to hierarchical masses among generations and it is implemented by choosing hierarchical entries in the $M_R={\rm diag}(M_{R_1},M_{R_2},M_{R_3})$ matrix. As for the $\mu_X={\rm diag}(\mu_{X_1},
\mu_{X_2},\mu_{X_3})$ matrix which introduces the tiny splitting within the heavy masses in the same generation we choose it here to be degenerate, $\mu_{X_{1,2,3}}=\mu_X$. We focus on the normal hierarchy $M_{R_1}<M_{R_2}<M_{R_3}$, since we have found similar conclusions for other hierarchies.

\begin{figure}[t!]
\begin{center}
\begin{tabular}{cc}
\hspace{-0.5cm}\includegraphics[width=.47\textwidth]{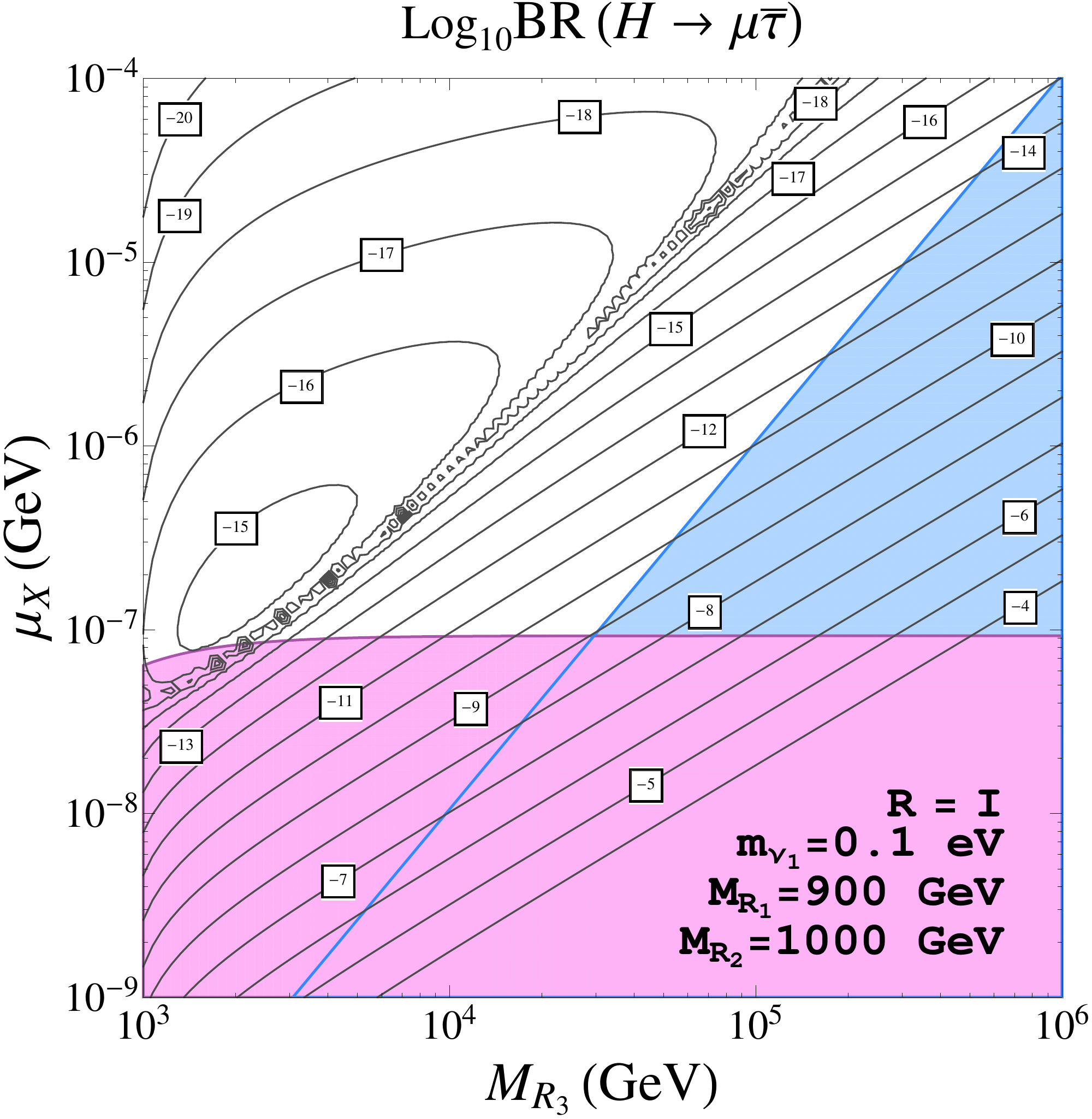} 
\end{tabular}
\caption{Contour lines of BR($H \to \mu \bar \tau$) in the $(M_{R_3},\mu_X)$ plane for $R=I$, $m_{\nu_1}=0.1\,{\rm eV}$, $M_{R_1}=900\,{\rm GeV}$ and $M_{R_2}=1000\,{\rm GeV}$. The shaded pink area is excluded by the upper bound on BR$(\mu \to e \gamma)$. The shaded blue area is excluded by the perturbativity requirement of the neutrino Yukawa couplings.}
\label{ContourPlot_MR3}
\end{center}
\end{figure}

Figure~\ref{ContourPlot_MR3} shows the contour lines of BR($H \to \mu \bar \tau$) in the $(M_{R_3},\mu_X)$ plane for $R=I$ and $m_{\nu_1}=0.1\,{\rm eV}$. It is clear from this contour plot that the behavior of the LFV rates in the hierarchical case with respect to the heaviest neutrino mass $M_{R_3}$ is very similar to the one found previously for the degenerate case with respect the common $M_{R}$. Again, there are dips in the BR($H \to \mu \bar \tau$) rates due to the destructive interferences among the contributing diagrams. We have found, for this hypothesis of degenerate $\mu_X$ and hierarchical $M_R$, an enhancement of the LFVHD rates by approximately one order of magnitude as compared to the degenerate case in most of the explored parameter space regions, where the maximum allowed BR$(H \to \mu \bar \tau)$ rates reach values up to about $10^{-9}$ for $M_{R_1}=900\,{\rm GeV}$, $M_{R_2}=1000\,{\rm GeV}$, $M_{R_3}=3\times 10^{4}\,{\rm GeV}$, $\mu_X=10^{-7}\,{\rm GeV}$, and $R=I$.

\section{LFV Rates in the Inverse Seesaw: Case II}
\label{resultsII}

In the line of thinking of increasing $H \to e \bar \tau$ and $H \to \mu \bar \tau$ rates, while reducing the predictions of $\mu \to e \gamma$ rates, which is the most restrictive constraint together with the Yukawa perturbativity, we propose in this section more general scenarios with non-diagonal $\mu_X$ and diagonal and degenerate $M_R$. In order to localize the class of scenarios leading to large and allowed LFVHD rates, we first make a rough estimate of the expected maximal rates for the $H \to \mu \bar \tau$ channel by using our approximate formula of Eq.~(\ref{FIThtaumu}) which is given just in terms of the neutrino Yukawa coupling
matrix $Y_\nu$ and $M_R$. The most constraining observables in our study are the LFV radiative decays, which limit the maximal value of the non-diagonal $(Y_\nu Y_\nu^\dagger)_{ij}$ entries. By using our approximate formula of Eq.~(\ref{approxformula}) and the present bounds in Eq.~(\ref{TAUMUGmax}), we obtain:
\begin{eqnarray}
v^2(Y_\nu Y_\nu^\dagger)_{12}^{\rm max}/M_R^2 & \sim & 2.5\times10^{-5},\label{12max}\\
v^2(Y_\nu Y_\nu^\dagger)_{13}^{\rm max}/M_R^2  & \sim  & 0.015,\label{13max}\\
v^2(Y_\nu Y_\nu^\dagger)_{23}^{\rm max}/M_R^2 & \sim  & 0.017\label{23max}.
\end{eqnarray}
Then, in order to simplify our search, and given the above relative strong suppression of the 12 element, it seems reasonable to neglect it against the other
off-diagonal elements. In that case, Eq.~(\ref{FIThtaumu}) for the $H \to \mu \bar \tau$ decay mode reads as
\begin{eqnarray}
{\rm BR}^{\rm approx}_{H\to\mu\bar\tau}&=&10^{-7}\frac{v^4}{M_R^4} \left| (Y_\nu Y_\nu^\dag)_{23} \right|^2 \\
&\times& \left| 1-5.7[(Y_\nu Y_\nu^\dag)_{22} + (Y_\nu Y_\nu^\dag)_{33}] \right|^2 \,. \nonumber
\end{eqnarray}
This equation clearly shows that the maximal BR($H\to\mu\bar\tau$) rates
are obtained for the maximum allowed values of $(Y_\nu Y_\nu^\dagger)_{23}$, $(Y_\nu Y_\nu^\dagger)_{22}$ and $(Y_\nu Y_\nu^\dagger)_{33}$. By setting the maximum allowed value for $v^2(Y_\nu Y_\nu^\dagger)_{23}^{\rm max}/M_R^2$ to that given in Eq.~(\ref{23max}) and fixing the values of $(Y_\nu Y_\nu^\dagger)_{22}$ and $(Y_\nu Y_\nu^\dagger)_{33}$ to their maximum values allowed by Eq.~(\ref{Yuk-pert}),
\begin{equation}
(Y_\nu Y_\nu^\dagger)_{33}^{\rm max}= (Y_\nu Y_\nu^\dagger)_{22}^{\rm max}= (Y_\nu Y_\nu^\dagger)_{11}^{\rm max}= 18 \pi \,,
\label{pert}
\end{equation}
we obtain our approximate prediction for the maximal rates:
\begin{equation}
 {\rm BR}^{\rm max}_{H\to\mu\bar\tau} \simeq 10^{-5} \,.
 \end{equation}
The same procedure can be applied to the $H \to e \bar \tau$ channel, leading to similar results on its maximal branching ratios. These maximal $H \to \mu \bar \tau$ and $H \to e \bar \tau$ rates are much more promising than the ones obtained in the previous section under the simple hypothesis of diagonal $\mu_X$ and $M_R,$ since they are closer to the expected LHC sensitivity for these channels (see for instance~\cite{Davidson:2012ds,Bressler:2014jta}) which will be surely improved in the forthcoming runs of the LHC.

Assuming this kind of Yukawa matrix which gives rise to the maximal BR($H \to \mu \bar \tau$) and BR($H \to e \bar \tau$), the issue now would be to find the possible $\mu_X$ matrices that keep the agreement with neutrino data. This fact will be always possible by means of Eq.~(\ref{muXtexture}) and because of the possibility of decoupling the low energy neutrino physics from the LFV physics in the ISS model.
This means that, for a given input $Y_\nu$, Eq.~(\ref{muXtexture}) tells us which $\mu_X$ keeps the agreement with neutrino data. We address the reader to our main article~\cite{Arganda:2014dta} for more details on this case II and for some illustrative examples of Yukawa textures producing these maximal LFVHD rates.

\section{Conclusions}
\label{conclu}
We have studied the LFVHD within the context of the ISS where three additional pairs of massive RH singlet neutrinos are added to the SM particle content.
The most relevant ISS parameters have been found to be $M_R$ and $Y_\nu$.
We have required that the input ISS parameters be compatible with the present neutrino data and other constraints, with the perturbativity of the neutrino Yukawa couplings and the present upper bounds on the LFV radiative decays being the most restrictive ones. 
The largest maximum LFVHD rates are BR($H \to e \bar \tau$) and BR($H \to \mu \bar \tau$) and under the hypothesis of diagonal $\mu_X$ and $M_R$ they can reach at the most values of $10^{-10}$ for the degenerate heavy neutrino case and $10^{-9}$ for the hierarchical case.
Going beyond this simple hypothesis, we find more general ISS scenarios with non-diagonal $\mu_X$ in which BR($\mu \to e \gamma$) is extremely suppressed and BR($H \to e \bar \tau$) or BR($H \to \mu \bar \tau$) are larger while respecting the BR($\tau \to e \gamma$) and BR($\tau \to \mu \gamma$) upper bounds. These rates reach maximal values of $10^{-5}$, being very promising for LFVHD searches at the LHC. 

\section*{Acknowledgments}

This work is supported by the European Union FP7 ITN
INVISIBLES (Marie Curie Actions, PITN- GA-2011- 289442), by the CICYT through the
project FPA2012-31880,  
by the Spanish Consolider-Ingenio 2010 Programme CPAN (CSD2007-00042) 
and by the Spanish MINECO's ``Centro de Excelencia Severo Ochoa'' Programme under grant SEV-2012-0249.
E.~A. is financially supported by the Spanish DGIID-DGA grant 2013-E24/2 and the Spanish MICINN grants FPA2012-35453 and CPAN-CSD2007-00042.
X.~M. is supported through the FPU grant AP-2012-6708.







\end{document}